\documentclass[prd,aps,showpacs, superscriptaddress,10pt]{revtex4-1}
\usepackage[utf8]{inputenc}
\usepackage{amsmath}
\usepackage{amssymb}
\usepackage{epsfig}
\usepackage{bbold}

\newcommand{\sgn}{\;\mbox{sgn}}
\newcommand{\e}{\mbox{e}}

\begin{document}

\title{Solutions to the problem of ELKO spinor localization in brane models}

\author{I. C. Jardim}  
\email{jardim@fisica.ufc.br} \affiliation{Departamento de F\'isica,
  Universidade Federal do Cear\'a, Caixa Postal 6030, Campus do Pici,
  60455-760 Fortaleza, Cear\'a, Brazil}

\author{G. Alencar}
\email{geova@fisica.ufc.br} \affiliation{Departamento de F\'isica,
 Universidade Federal do Cear\'a, Caixa Postal 6030, Campus do Pici,
 60455-760 Fortaleza, Cear\'a, Brazil}
 
 \author{R. R. Landim}  
\email{renan@fisica.ufc.br} \affiliation{Departamento de F\'isica,
  Universidade Federal do Cear\'a, Caixa Postal 6030, Campus do Pici,
  60455-760 Fortaleza, Cear\'a, Brazil}

 \author{R. N. Costa Filho} 
\email{rai@fisica.ufc.br} \affiliation{Departamento de F\'isica,
 Universidade Federal do Cear\'a, Caixa Postal 6030, Campus do Pici,
  60455-760 Fortaleza, Cear\'a, Brazil}

 \begin{abstract}
In this paper we present two different solutions to the problem of zero mode localization of ELKO spinor. In a recent paper the present authors reopened this problem since the solution presented before did not satisfy the boundary condition at the origin. The first solution is given by the introduction of a mass term and by coupling the spinor with the brane through a delta function. The second solution is reached by a Yukawa geometrical coupling with the Ricci scalar. This two models changes consistently the the boundary condition at infinity and at the origin. For the case of Geometrical coupling we are able to show that the zero mode is localized for any smooth version of the RS model.
\end{abstract}

\maketitle
\section{Introduction}
The a idea that space-time has more than four dimension has drawn some attention of the community in the past after the idea of Kaluza and Klein about compact extra dimensions. The main idea is that the extra dimension are so tiny that can not be observed and in this way the "escaping" of the fields to the extra dimensions becomes a small correction \cite{Bailin:1987jd}. Depending on the compactification and the number of dimensions different kinds of fields in the lower dimensional theory can be obtained\cite{Salam:1981xd}. In fact these models provide us with a plethora of massive states, the zero mode being just a particular case. 

In the late 90 models considering the universe as a branes in a higher dimensional gained much attention again \cite{Polchinski:1995mt}. A scenario considering our world as a shell has been proposed in \cite{Gogberashvili:1998iu} and further developed in \cite{Jardim:2011gg,Jardim:2013jy,Akama:2011wi}. Probably based in \cite{Gogberashvili:1998iu} Lisa Randall and Raman Sundrum (RS) proposed another scenario with four dimensional branes in a five dimensional space with negative cosmological constant. In this scenario two different models has been considered: In the type I RS  a compact space with two branes and a $Z_2$ symmetry is used, solving the hierarchy problem \cite{Randall:1999ee}. Being a model with compact dimension the dimensional reduction works in a very similar way to the KK model; In the type II RS model just one brane embedded in a large extra dimension space is considered. The extra dimension is curved by a warp factor such that the model has been considered as an alternative to 
compactification \cite{Randall:1999vf}.  As a model for large extra dimensions the issue of zero mode localization of fields is important. In fact in the last ten years the zero mode localization of gauge field has becomes a drawn back in these model. This localization is necessary since in a four dimension space fields propagating into the bulk can not be observed. Moreover, it has been found that the zero mode of gravity and scalar fields are localized \cite{Randall:1999vf,Bajc:1999mh} in a positive tension brane. However, due to its conformal invariance the vector field is not localized, which is a serious problem for a realistic model. 

The solution to the above problem has been studied in many ways. For example,  some authors introduced a dilaton coupling to solve the problem \cite{Kehagias:2000au}, and others like in \cite{Dvali:1996xe} proposed that a strongly coupled gauge theory in five dimensions can generate a massless photon in the brane. Generalization considering anti-symmetric fields can also be found in the literature \cite{Kaloper:2000xa,Duff:1980qv,Duff:2000se,Alencar:2010mi,Alencar:2010hs,Alencar:2010vk,Landim:2010pq,Fu:2012sa}. 
Most of these models introduces other fields or nonlinearities to the gauge field \cite{Chumbes:2011zt}. However there has been two proposes that do not introduces new fields neither nonlinearities. One is by the introduction of a mass term and a delta interaction with the brane \cite{Ghoroku:2001zu}. This kind of model has been generalized to consider smooth branes and to $p-$forms by the present authors in \cite{Jardim:2014vba}. The other is a very recent model in which the gauge field is coupled to the Ricci scalar, what was called a geometrical coupling \cite{Alencar:2014moa,Alencar:2014fga}.  Similar ideas has been used before but by coupling with the field strength \cite{Germani:2011cv}. 
By another hand, an interesting viewpoint is related to models where membranes are smoothed out by topological defects \cite{Bazeia:2005hu,Landim:2011ki,Landim:2011ts,Alencar:2012en}. The advantage of these models is that the $\delta$-function singularities generated by the brane in the RS scenario are eliminated. This kind of generalization also provides methods for finding analytical solutions \cite{Cvetic:2008gu,Landim:2013dja}. 

Studies of matter fields with half spin has also been considered for cases with and without an interaction term, where localization and resonances are studied \cite{Liu:2009ve,Zhao:2009ja,Liang:2009zzf,Zhao:2010mk,Zhao:2011hg,Li:2010dy,
Castro:2010au,Correa:2010zg,Castro:2010uj,Chumbes:2010xg,Castro:2011pp}.
Another kind of fields that can be considered is the ELKO spinor. This field has spin 1/2, mass dimension one and can be considered as a first-principle candidate of dark matter \cite{Ahluwalia:2004sz,Ahluwalia:2004ab,Gillard:2009zw,Ahluwalia:2010zn,Ahluwalia:2008xi}. For more explanations and references see for example Ref. \cite{daRocha:2011yr}. In the framework of braneworld scenarios, the localization
of  ELKO spinors on branes has been considered in Ref. \cite{Liu:2011nb}. However the present authors showed very recently in Ref. \cite{Jardim:2014cya} that the affirmation of the last authors that the zero mode of this field is  localized is equivocated. This reopened the problem of the localization of ELKO spinor. Here we show that the previous proposes of the present authors to solve the problem of $p-$form field localization in Refs. \cite{Jardim:2014vba,Alencar:2014moa}  can also be used to this problem. We show that for specific values of the coupling constant we can have localized zero modes for the ELKO spinor. This rises the same and interesting question about the origin of this kind of coupling.

This paper is organized as follows: second section gives a review about the way to obtain the mass equation for the ELKO spinors. Third section is devoted to study the solution to the problem by using a mass term and a cubic interaction with the delta function. In the fourth section we use the geometrical coupling as another solution the the localization of the zero mode. Finally in the fifth section we discuss the conclusions and perspectives.

\section{Review of localization of 5D ELKO spinors with coupling term on Minkowski brane}
In this section we will make a review of localization of $5D$ ELKO spinor, based in \cite{Liu:2011nb}. The action used in the cited article for ELKO spinor is
\begin{equation}\label{action5}
S = \int d^{5}x\sqrt{-g}\left[-\frac{1}{4}\left(D_{M}\lambda D^{M}\bar{\lambda} + D_{M}\bar{\lambda} D^{M}\lambda\right) -\eta F(z)\bar{\lambda}\lambda\right] ,
\end{equation}
where $\eta$ in a coupling constant, $F(z)$ is a scalar function of conformal extra dimension $z$ and $D_{M}$ is the covariant derivative defined as
\begin{equation}
 D_{M}\lambda = \partial_{M}\lambda +\Omega_{M}\lambda.
\end{equation}
As show in \cite{Liu:2011nb} the nonvanish components of spin connection are
\begin{equation}
 \Omega_{\mu} = \frac{1}{2}A'(z)\gamma_{\mu}\gamma_{5},
\end{equation}
where primes denotes derivative with respect to the argument, $A(z)$ is the conformal warp factor $g_{MN} =\e^{2A(z)}\eta_{MN}$ and $\eta_{MN} = \mbox{Diag.}(-1,1,1,1,1)$.  
Taking the variation of action in respect to $\bar{\lambda}$ we obtain the equation of motion 
\begin{equation}
 D_{M}\left[\sqrt{-g}D^{M}\lambda\right] -2\eta\sqrt{-g}F(z)\lambda = 0,
\end{equation}
using the metric and the nonvanish components of spin connection, we can write the above equation in the form 
\begin{eqnarray}\label{lambda}
 &&\eta^{\mu\nu}\partial_{\mu}\partial_{\nu}\lambda -A'(z)\gamma_{5}\eta^{\mu\nu}\gamma_{\mu}\partial_{\nu}\lambda  -A'^{2}\lambda + \e^{-3A}(\e^{3A}\lambda')' -2\eta\e^{2A}F(z)\lambda = 0
\end{eqnarray}
Due the term $A'(z)\gamma_{5}\eta^{\mu\nu}\gamma_{\mu}\partial_{\nu}\lambda $ the authors of \cite{Liu:2011nb} propose a decomposition of ELKO field as $\lambda = \lambda_{+} +\lambda _{-}$, where  
\begin{equation}
 \lambda_{\pm} = \e^{-3A/2}\sum_{n}\alpha_{n}(z)\tilde{\lambda}_{\pm}^{n}(x)= \e^{-3A/2}\sum_{n}\alpha_{n}(z)\left[\varsigma_{\pm}^{n}(x) +\tau_{\pm}^{n}(x)\right].
\end{equation}
 $\varsigma_{\pm}^{n}(x)$ and $\tau_{\pm}^{n}(x)$ are two independent $4D$ ELKO field which satisfy the Klein-Gordon equations $\Box\tau_{\pm}^{n}(x) = m_{n}^{2}\tau_{\pm}^{n}(x)$, $\Box\varsigma_{\pm}^{n}(x) = m_{n}^{2}\varsigma_{\pm}^{n}(x)$ and 
 the relations $\gamma^{5}\tau_{\pm} = \mp\varsigma_{\pm}$, $\gamma^{5}\varsigma_{\pm} = \pm\tau_{\pm}$. This decomposition in eq. (\ref{lambda}) leads to
 \begin{equation}\label{alphaF}
\alpha_{n}''(z)-\left(\frac{13A'^{2}}{4} +\frac{3A''}{2} -m_{n}^{2}+im_{n}A'(z)+2\eta\e^{2A}F(z)\right)\alpha_{n}(z)=0,
\end{equation}
with the normalization condition
\begin{equation}
 \int \alpha_{n}^{*}\alpha_{m}dz = \delta_{mn}.
\end{equation}
The eq. (\ref{alphaF}) is the general equation of localization coefficients $\alpha_{n}$. In following sections we will compute this coefficients for some functions $F(z)$.  

\section{Localization of 5D ELKO spinors with  $4D$ and $5D$ mass term}
In this section we will compute explicitly the solution of eq. (\ref{alphaF}) in case that $F(z)$ is a $5D$ mass and a $4D$ mass term, i.e.
\begin{equation}
 \eta F(z) = \frac{1}{2}\left(M^{2} +c\delta(z)\right).
\end{equation}
These kind of coupling has used for $p$-form field and provide an efficient mechanism to trap the zero modes \cite{Jardim:2014vba}.
In this case the equation of $\alpha_{n}$ can be written as
\begin{equation}\label{alphaMc}
\alpha_{n}''(z)-\left[\left(\frac{19k^{2}}{4} +M^{2}\right)[k|z| +1]^{-2} +(c-3k)\delta(z) -m_{n}^{2}-im_{n}k\sgn(z)[k|z| +1]^{-1}\right]\alpha_{n}(z)=0,
\end{equation}
where we used the conformal Randall-Sundrum warp factor
\begin{equation}
 A(z) = -\ln[k|z| +1].
\end{equation}
For the zero mode, $m_{0} = 0$, the eq. (\ref{alphaMc}) provides the solution
\begin{equation}
 a_{0} = C_{+}(k|z| +1)^{1/2 +\nu} +C_{-}(k|z| +1)^{1/2 -\nu} 
\end{equation}
where  $C_{+}$ and $ C_{-}$ are constants and $\nu$ is given by
\begin{equation}\label{nuMc}
 \nu =  \sqrt{5 +M^{2}/k^{2}}.
\end{equation}
The boundary condition at origin imposes the relation between the constants
\begin{equation}
 \left(2k(2 +\nu)-c\right)C_{+} +\left(2k(2-\nu)-c\right)C_{-}  =0.
\end{equation}
To obtain a convergent solution we must to fix
\begin{equation}\label{cMc}
 c = 2k(2-\nu).
\end{equation}
This procedure leads to the normalized convergent solution
\begin{equation}
 a_{0} = \frac{(k|z| +1)^{1/2 -\nu}}{\sqrt{k(\nu-1)}}.
\end{equation}
As we can see from eqs. (\ref{nuMc}) and (\ref{cMc}) it is not possible vanish $M$ and $c$ at same time and keep the solution convergent. In this point the ref. \cite{Liu:2011nb} is incomplete, because its not satisfy the boundary condition at $z=0$. 
\\ For massive modes the field eq. (\ref{alphaMc}) can be written in the form
\begin{equation}
a_{n}''(z)-\left[\left(\frac{19k^{2}}{4} +M^{2}\right)[k|z| +1]^{-2} +(c-3k)\delta(z) -m_{n}^{2}-im_{n}k\sgn(z)[k|z| +1]^{-1}\right]a_{n}(z)=0.
\end{equation}
The above equation has as solution 
 \begin{eqnarray}
 a_{n}(z) &=& C_{1}\theta(z)M_{1/2,\nu}\left(i2m_{n}/k(k|z|+1)\right) +C_{2}\theta(-z)M_{-1/2,\nu}\left(i2m_{n}/k(k|z|+1)\right) + \nonumber
 \\&&+C_{3}\theta(z)W_{1/2,\nu}\left(i2m_{n}/k(k|z|+1)\right) + C_{4}\theta(-z)W_{-1/2,\nu}\left(i2m_{n}/k(k|z|+1)\right),
\end{eqnarray}
where $C_{1},C_{2} ,C_{3}$ e $C_{2}$ are constants which must satisfy the following boundary conditions
\begin{eqnarray}
&& C_{1}M_{1/2,\nu}\left(2im_{n}/k\right) -C_{2}M_{-1/2,\nu}\left(2im_{n}/k\right) + C_{3}W_{1/2,\nu}\left(2im_{n}/k\right) -C_{4}W_{-1/2,\nu}\left(2im_{n}/k\right) =0 
\\&&  4im_{n}\left[ C_{1}M'_{1/2,\nu}\left(u\right) +C_{2}M'_{-1/2,\nu}\left(u\right)\right] -(c-3k)( C_{1}M_{1/2,\nu}\left(im_{n}/k\right) +C_{2}M_{-1/2,\nu}\left(im_{n}/k\right) ) + \nonumber
\\&& + 4im_{n}\left[ C_{3}W'_{1/2,\nu}\left(u\right) +C_{4}W'_{-1/2,\nu}\left(u\right)\right] -(c-3k)( C_{3}W_{1/2,\nu}\left(im_{n}/k\right) +C_{4}W_{-1/2,\nu}\left(im_{n}/k\right) ) =0.
\end{eqnarray}
At this point could be interesting localize a specific massive mode to model the dark matter by ELKO spinor without the Higgs mechanism, but due the argument of Whittaker function be complex it is not possible to find a coupling constant $c$ which localize any massive mode. So, to keep the ELKO spinor as a candidate to dark mater the Higgs mechanism is still necessary.

\section{ELKO Spinor with geometrical coupling}
In this section we will use the geometrical coupling, which consist in a Yukawa interaction with the Ricci scalar. This kind of coupling has used for $p$-form and provides a second efficient method to localize the zero modes \cite{Alencar:2014moa, Alencar:2014fga}. In RS brane it provides an natural source for the coupling term used in previous section. The action is given by
\begin{equation}\label{SR5}
S = \int d^{5}x\sqrt{-g}\left[-\frac{1}{4}\left(D_{M}\lambda D^{M}\bar{\lambda} + D_{M}\bar{\lambda} D^{M}\lambda\right) -\eta R\lambda\bar{\lambda}\right],
\end{equation}
i.e., is equivalent to make $F(z) = R$ in (\ref{action5}). In conformal coordinates the Ricci scalar is given by
\begin{equation}
  R = -4(2A'' +3A'^{2})\e^{-2A},
\end{equation}
and the field equation (\ref{alphaF}) can be written as
\begin{equation}
a_{n}''(z)-\left[\left(\frac{13}{4}-24\eta\right)A'^{2} +\left(\frac{3}{2}-16\eta\right)A'' -m_{n}^{2}+im_{n}A'(z)\right]a_{n}(z)=0.
\end{equation}
Like in refs. \cite{Alencar:2014moa, Alencar:2014fga} we will propose a solution for zero mode in the form 
\begin{equation}\label{a0R}
 a_{0} \propto \e^{\gamma A},
\end{equation}
so that the field equation provides the condition for this kind of solution
\begin{equation}\label{cgamma}
\left(\gamma^{2}-\frac{13}{4}+24\eta\right)A'^{2} +\left(\gamma-\frac{3}{2}+16\eta\right)A''=0.
\end{equation}
As we can see form (\ref{a0R}), to obtain a convergent solution is necessary that $\gamma > 0$. With this restriction the eq. (\ref{cgamma}) provides as solution $\gamma = 2$ e $\eta = -1/32$. Therefore in this model the localization of the zero mode is guaranteed for any smooth version of RS model. This solves the problem of zero mode localization of ELKO spinors.
To obtain explicit solutions, as well as the massive modes, we must to use specific scenarios, i.e., explicit warp factors. This is what we will compute in following subsections.

\subsection{Randall-Sundrum brane}
The first case that we will compute the explicit solution is the Randall-Sundrum brane scenario. In this case the warp factor in a conformal coordinate is given by
\begin{equation}
 A(z) = -\ln[k|z|+1].
\end{equation}
Replacing this warp factor in  (\ref{a0R}) we obtain the normalized convergent solution for zero mode,
\begin{equation}
  a_{0} =\sqrt{\frac{2}{3k}} (k|z|+1)^{-2}.
\end{equation}
Comparing  with the solution obtained in Sec. III, we conclude that the geometrical coupling is equivalent to make $M^{2}/k^{2} = 5/4$, and from (\ref{cMc}), $c = -k$.
This correspondence occur due the fact that the geometrical coupling in a Randall-Sundrum case reduces to
\begin{equation}
 \eta R\lambda\bar{\lambda} = \frac{1}{2}\left(\frac{5}{4}k^{2} -k\delta(z)\right)\lambda\bar{\lambda}, 
\end{equation}
i.e., an equivalent term used in Sec. III. Due this equivalence the massive modes is given by
\begin{eqnarray}
 a_{n}(z) &=& \theta(z)\left[ A_{1}M_{1/2,5/2}\left(i2m_{n}/k(k|z|+1)\right) +B_{1}W_{1/2,5/2}\left(i2m_{n}/k(k|z|+1)\right)\right] + \nonumber
\\&& +  \theta(-z)\left[ A_{2}M_{-1/2,5/2}\left(i2m_{n}/k(k|z|+1)\right) +B_{2}W_{-1/2,5/2}\left(i2m_{n}/k(k|z|+1)\right)\right],
 \end{eqnarray}
where $A_{1},A_{2} ,B_{1}$ e $B_{2}$ are constants which must satisfy the following conditions
\begin{eqnarray}
&& A_{1}M_{1/2,5/2}\left(u_{0}\right) -A_{2}M_{-1/2,5/2}\left(u_{0}\right) + B_{1}W_{1/2,5/2}\left(u_{0}\right) -B_{2}W_{-1/2,5/2}\left(u_{0}\right) =0 \nonumber
\\&& u_{0}\biggl.\left[ A_{1}M'_{1/2,5/2}\left(u\right) +A_{2}M'_{-1/2,5/2}\left(u\right)\right]\biggl|_{u= u_{0}} +2\left[ A_{1}M_{1/2,5/2}\left(u_{0}\right) +A_{2}M_{-1/2,5/2}\left(u_{0}\right) \right] + \nonumber
\\&& + u_{0}\biggl.\left[ B_{1}W'_{1/2,5/2}\left(u\right) +B_{2}W'_{-1/2,5/2}\left(u\right)\right]\biggl|_{u= u_{0}} +2\left[ B_{1}W_{1/2,5/2}\left(u_{0}\right) +B_{2}W_{-1/2,5/2}\left(u_{0}\right) \right] =0
\end{eqnarray}
where $u_{0} = 2im_{n}/k$. Due the asymptotic behavior of Whittaker functions with complex argument is not possible to find a convergent solutions, i.e., the massive modes are non-localized.  
\subsection{Smooth brane}
In this section we will use the following smooth warp factor \cite{Du:2013bx, Melfo:2002wd}
\begin{equation}
  A(z) = -\frac{1}{2n}\ln\left[\left(kz\right)^{2n}+1\right],
\end{equation}
which recover the Randall-Sundrum metric at large $z$ for $n$ integer. Replacing this warp factor in (\ref{a0R}) we find the normalized convergent solution for zero mode
\begin{equation}
  a_{0} = \sqrt{\frac{k}{2}\frac{\Gamma(2/n)}{\Gamma(3/2n)\Gamma(1+1/2n)}} [(kz)^{2n}+1]^{-1/n}.
\end{equation}
In massive case the field equation can be written as
\begin{equation}
a_{n}''(z)-\left[6\frac{(kz)^{2n-2}k^{2}}{[(kz)^{2n}+1]} -4(n+1)\frac{(kz)^{2n-2}k^{2}}{[(kz)^{2n}+1]^{2}} -m_{n}^{2} -im_{n}\frac{(kz)^{2n-1}k}{[(kz)^{2n}+1]}\right]a_{n}(z)=0. 
\end{equation}
The solution of massive modes can not be find analytically, but since the asymptotic behavior is the same of Randall-Sundrum case we can conclude that it is not possible to find a convergent solution.
Due the imaginary massive term is not possible to use the transference matrix method to study the unstable massive modes. 
\section {Conclusion}
In this paper we used two different methods to localize the zero mode of ELKO spinor. This problem has been reopened since the results of Ref. \cite{Liu:2011nb} was shown to be equivocated in Ref. \cite{Jardim:2014cya}. The fist method discussed in this work is the inclusion of  $4D$ and $5D$ mass terms. We find the
relation between this two mass terms which localize the zero mode. This result show that it is not possible to vanish booth contributions for a localized solution, in disagreement with the results obtained in \cite{Liu:2011nb}.
This disagreement occur because the authors of cited article find a convergent solution but they do not take into account the boundary condition at the origin. We also show that the massive modes are non-localized for this new model.
The reason is that the solution is given by Whittaker function with complex argument, so it is not possible to find a coupling constant $c$ which localize an specific massive mode to model
the dark matter by ELKO spinor without the Higgs mechanism.

The other method used to localize the zero mode of ELKO spinor is the geometrical coupling. In this model we compute the coupling constant that localize the zero mode for all warp factor
which has Randall-Sundrum asymptotic behavior. We compute the explicit solution for Randall-Sundrum case and for a specific smooth brane scenario. Like the previous mechanism the massive modes are non-localized. Due to the complex term in the potential the authors has not been able to compute resonances and this is left to future work.
\section*{Acknowledgments}

We acknowledge the financial support provided by Funda\c c\~ao Cearense de Apoio ao Desenvolvimento Cient\'\i fico e Tecnol\'ogico (FUNCAP), the Conselho Nacional de 
Desenvolvimento Cient\'\i fico e Tecnol\'ogico (CNPq) and FUNCAP/CNPq/PRONEX.


\begin{thebibliography}{99}
\bibitem{Bailin:1987jd} 
  D.~Bailin and A.~Love,
  Rept.\ Prog.\ Phys.\  {\bf 50}, 1087 (1987).
 
\bibitem{Salam:1981xd} 
  A.~Salam and J.~A.~Strathdee,
  Annals Phys.\  {\bf 141}, 316 (1982).
  
\bibitem{Polchinski:1995mt} 
  J.~Polchinski,
  Phys.\ Rev.\ Lett.\  {\bf 75}, 4724 (1995)
  [hep-th/9510017].
  
\bibitem{Gogberashvili:1998iu} 
  M.~Gogberashvili,
  Europhys.\ Lett.\  {\bf 49}, 396 (2000)
  [hep-ph/9812365].
  
\bibitem{Jardim:2011gg} 
  I.~C.~Jardim, R.~R.~Landim, G.~Alencar and R.~N.~Costa Filho,
  Phys.\ Rev.\ D {\bf 84}, 064019 (2011)
  [arXiv:1105.4578 [gr-qc]].
  
\bibitem{Jardim:2013jy} 
  I.~C.~Jardim, R.~R.~Landim, G.~Alencar and R.~N.~Costa Filho,
  Phys.\ Rev.\ D {\bf 88}, no. 2, 024004 (2013)
  [arXiv:1301.2578 [gr-qc]].
  
\bibitem{Akama:2011wi} 
  K.~Akama, T.~Hattori and H.~Mukaida,
  arXiv:1109.0840 [gr-qc].
  
\bibitem{Randall:1999ee} 
  L.~Randall and R.~Sundrum,
  Phys.\ Rev.\ Lett.\  {\bf 83}, 3370 (1999)
  [hep-ph/9905221].
  
    
 \bibitem{Randall:1999vf}
Lisa Randall and Raman Sundrum,
  \href{http://dx.doi.org/10.1103/PhysRevLett.83.4690}{{\em Phys.Rev.Lett.}
  {\bfseries 83} (1999) 4690--4693},
\href{http://arxiv.org/abs/hep-th/9906064}{{\ttfamily arXiv:hep-th/9906064
  [hep-th]}}.

\bibitem{Bajc:1999mh} 
  B.~Bajc and G.~Gabadadze,
  Phys.\ Lett.\ B {\bf 474}, 282 (2000)
  [hep-th/9912232].
  

\bibitem{Kehagias:2000au} 
  A.~Kehagias and K.~Tamvakis,
  Phys.\ Lett.\ B {\bf 504}, 38 (2001)
  [hep-th/0010112].

\bibitem{Dvali:1996xe} 
  G.~R.~Dvali and M.~A.~Shifman,
  Phys.\ Lett.\ B {\bf 396}, 64 (1997)
  [Erratum-ibid.\ B {\bf 407}, 452 (1997)]
  [hep-th/9612128].
  
\bibitem{Kaloper:2000xa} 
  N.~Kaloper, E.~Silverstein and L.~Susskind,
  JHEP {\bf 0105}, 031 (2001)
  [hep-th/0006192].
  
\bibitem{Duff:1980qv} 
  M.~J.~Duff and P.~van Nieuwenhuizen,
  Phys.\ Lett.\ B {\bf 94}, 179 (1980).
  
\bibitem{Duff:2000se} 
  M.~J.~Duff and J.~T.~Liu,
  Phys.\ Lett.\ B {\bf 508}, 381 (2001)
  [hep-th/0010171].

\bibitem{Alencar:2010mi} 
  G.~Alencar, M.~O.~Tahim, R.~R.~Landim, C.~R.~Muniz and R.~N.~Costa Filho,
  Phys.\ Rev.\ D {\bf 82}, 104053 (2010)
  [arXiv:1005.1691 [hep-th]].
  
\bibitem{Alencar:2010hs} 
  G.~Alencar, R.~R.~Landim, M.~O.~Tahim, C.~R.~Muniz and R.~N.~Costa Filho,
  Phys.\ Lett.\ B {\bf 693}, 503 (2010)
  [arXiv:1008.0678 [hep-th]].
  
\bibitem{Alencar:2010vk} 
  G.~Alencar, R.~R.~Landim, M.~O.~Tahim, K.~C.~Mendes, R.~R.~Landim, M.~O.~Tahim, R.~N.~C.~Filho and K.~C.~Mendes,
  Europhys.\ Lett.\  {\bf 93}, 10003 (2011)
  [arXiv:1009.1183 [hep-th]].
 
\bibitem{Landim:2010pq} 
  R.~R.~Landim, G.~Alencar, M.~O.~Tahim, M.~A.~M.~Gomes and R.~N.~Costa Filho,
  Europhys.\ Lett.\  {\bf 97}, 20003 (2012)
  [arXiv:1010.1548 [hep-th]].
   
 

\bibitem{Fu:2012sa} 
  C.~E.~Fu, Y.~X.~Liu, K.~Yang and S.~W.~Wei,
  JHEP {\bf 1210}, 060 (2012)
  [arXiv:1207.3152 [hep-th]].
  
  
\bibitem{Chumbes:2011zt} 
  A.~E.~R.~Chumbes, J.~M.~Hoff da Silva and M.~B.~Hott,
  Phys.\ Rev.\ D {\bf 85}, 085003 (2012)
  [arXiv:1108.3821 [hep-th]].
 
\bibitem{Ghoroku:2001zu} 
  K.~Ghoroku and A.~Nakamura,
  Phys.\ Rev.\ D {\bf 65}, 084017 (2002)
  [hep-th/0106145].

\bibitem{Jardim:2014vba} 
  I.~C.~Jardim, G.~Alencar, R.~R.~Landim and R.~N.~C.~Filho,
  arXiv:1410.6756 [hep-th].

\bibitem{Alencar:2014moa} 
  G.~Alencar, R.~R.~Landim, M.~O.~Tahim and R.~N.~C.~Filho,
  arXiv:1409.4396 [hep-th].
  
\bibitem{Alencar:2014fga} 
  G.~Alencar, R.~R.~Landim, M.~O.~Tahim and R.~N.~C.~Filho,
  arXiv:1409.5042 [hep-th].
  

  
\bibitem{Germani:2011cv} 
  C.~Germani,
  Phys.\ Rev.\ D {\bf 85}, 055025 (2012)
  [arXiv:1109.3718 [hep-ph]].
  

   \bibitem{Bazeia:2005hu}
D.~Bazeia and L.~Losano,
  \href{http://dx.doi.org/10.1103/PhysRevD.73.025016}{{\em Phys.Rev.}
  {\bfseries D73} (2006) 025016},
\href{http://arxiv.org/abs/hep-th/0511193}{{\ttfamily arXiv:hep-th/0511193
  [hep-th]}}.




\bibitem{Landim:2011ki} 
  R.~R.~Landim, G.~Alencar, M.~O.~Tahim and R.~N.~Costa Filho,
  JHEP {\bf 1108}, 071 (2011)
  [arXiv:1105.5573 [hep-th]].
  
\bibitem{Landim:2011ts} 
  R.~R.~Landim, G.~Alencar, M.~O.~Tahim and R.~N.~Costa Filho,
  JHEP {\bf 1202}, 073 (2012)
  [arXiv:1110.5855 [hep-th]].
  
\bibitem{Alencar:2012en} 
  G.~Alencar, R.~R.~Landim, M.~O.~Tahim and R.~N.~C.~Filho,
  JHEP {\bf 1301}, 050 (2013)
  [arXiv:1207.3054 [hep-th]].

\bibitem{Cvetic:2008gu}
Mirjam Cvetic and Marko Robnik,
  \href{http://dx.doi.org/10.1103/PhysRevD.77.124003}{{\em Phys.Rev.}
  {\bfseries D77} (2008) 124003},
\href{http://arxiv.org/abs/0801.0801}{{\ttfamily arXiv:0801.0801 [hep-th]}}.

\bibitem{Landim:2013dja} 
  R.~R.~Landim, G.~Alencar, M.~O.~Tahim and R.~N.~Costa Filho,
  Phys.\ Lett.\ B {\bf 731}, 131 (2014)
  [arXiv:1310.2147 [hep-th]].
  
  
\bibitem{Liu:2009ve}
  Y.~X.~Liu, J.~Yang, Z.~H.~Zhao, C.~E.~Fu and Y.~S.~Duan,
  ``Fermion Localization and Resonances on A de Sitter Thick Brane,''
  Phys.\ Rev.\  D {\bf 80}, 065019 (2009)
  [arXiv:0904.1785 [hep-th]].


\bibitem{Zhao:2009ja}
  Z.~H.~Zhao, Y.~X.~Liu and H.~T.~Li,
  ``Fermion localization on asymmetric two-field thick branes,''
  Class.\ Quant.\ Grav.\  {\bf 27}, 185001 (2010)
  [arXiv:0911.2572 [hep-th]].

\bibitem{Liang:2009zzf}
  J.~Liang and Y.~S.~Duan,
  ``Localization of matter and fermion resonances on double walls,''
  Phys.\ Lett.\  B {\bf 681}, 172 (2009).

\bibitem{Zhao:2010mk}
  Z.~H.~Zhao, Y.~X.~Liu, H.~T.~Li and Y.~Q.~Wang,
  ``Effects of the variation of mass on fermion localization and resonances on
  thick branes,''
  Phys.\ Rev.\  D {\bf 82}, 084030 (2010)
  [arXiv:1004.2181 [hep-th]].

\bibitem{Zhao:2011hg}
  Z.~H.~Zhao, Y.~X.~Liu, Y.~Q.~Wang and H.~T.~Li,
  arXiv:1102.4894 [hep-th].

\bibitem{Li:2010dy}
  H.~T.~Li, Y.~X.~Liu, Z.~H.~Zhao and H.~Guo,
  Phys.\ Rev.\  D {\bf 83}, 045006 (2011)
  [arXiv:1006.4240 [hep-th]].

\bibitem{Castro:2010au}
  L.~B.~Castro,
  ``Fermion localization on two-field thick branes,''
  Phys.\ Rev.\  D {\bf 83}, 045002 (2011)
  [arXiv:1008.3665 [hep-th]].

\bibitem{Correa:2010zg}
  R.~A.~C.~Correa, A.~de Souza Dutra and M.~B.~Hott,
  ``Fermion localization on degenerate and critical branes,''
  arXiv:1011.1849 [hep-th].

\bibitem{Castro:2010uj}
  L.~B.~Castro and L.~A.~Meza,
  ``Fermion localization on branes with generalized dynamics,''
  arXiv:1011.5872 [hep-th].

\bibitem{Chumbes:2010xg}
  A.~E.~R.~Chumbes, A.~E.~O.~Vasquez and M.~B.~Hott,
  ``Fermion localization on a split brane,''
  arXiv:1012.1480 [hep-th].


\bibitem{Castro:2011pp}
  L.~B.~Castro and L.~A.~Meza,
  ``Effect of the variation of mass on fermion localization on thick branes,''
  arXiv:1104.5402 [hep-th].

  
\bibitem{Ahluwalia:2004sz}
  D.~V.~Ahluwalia-Khalilova and D.~Grumiller,
   Phys. Rev. D. {\bf72}, 067701 (2005),
  arXiv:hep-th/0410192.

\bibitem{Ahluwalia:2004ab}
  D.~V.~Ahluwalia-Khalilova and D.~Grumiller,
  JCAP {\bf 0507}, 012 (2005),
  arXiv:hep-th/0412080.

\bibitem{Gillard:2009zw}
  A.~B.~Gillard and B.~M.~S.~Martin,
  {\it{Proceedings of 7th international Heidelberg Conference on Dark 2009}},
  {\it{Christchurch, New Zealand, 18 24 January 2009}}.
  arXiv:0904.2063 [hep-th].

\bibitem{Ahluwalia:2010zn}
  D.~V.~Ahluwalia-Khalilova and S.~P.~Horvath,
  JHEP {\bf 1011}, 078 (2010),
  arXiv:1008.0436 [hep-ph].

\bibitem{Ahluwalia:2008xi}
  D.~V.~Ahluwalia-Khalilova, C.~-Y.~Lee, D.~Schritt and T.~F.~Watson,
  Phys.\ Lett.\ B {\bf 687}, 248 (2010),
  arXiv:0804.1854 [hep-th].
  
\bibitem{daRocha:2011yr} 
  R.~da Rocha, A.~E.~Bernardini and J.~M.~Hoff da Silva,
  JHEP {\bf 1104}, 110 (2011)
  [arXiv:1103.4759 [hep-th]].

\bibitem{Liu:2011nb} 
  Y.~X.~Liu, X.~N.~Zhou, K.~Yang and F.~W.~Chen,
  Phys.\ Rev.\ D {\bf 86}, 064012 (2012)
  [arXiv:1107.2506 [hep-th]].
 
\bibitem{Jardim:2014cya} 
  I.~C.~Jardim, G.~Alencar, R.~R.~Landim and R.~N.~C.~Filho,
  arXiv:1411.5980 [hep-th].
  
\bibitem{Du:2013bx} 
  Y.~Z.~Du, L.~Zhao, Y.~Zhong, C.~E.~Fu and H.~Guo,
  Phys.\ Rev.\ D {\bf 88}, 024009 (2013)
  [arXiv:1301.3204 [hep-th], arXiv:1301.3204 [hep-th]].

\bibitem{Melfo:2002wd} 
  A.~Melfo, N.~Pantoja and A.~Skirzewski,
  Phys.\ Rev.\ D {\bf 67}, 105003 (2003)
  [gr-qc/0211081].
  
  
\end{thebibliography}
\end{document}